\documentstyle[12pt]{article}
\begin{document}
\begin{center}

{\large \bf SEMICLASSICAL AND QUANTUM FIELD THEORETIC BOUNDS FOR
 TRAVERSABLE  LORENTZIAN  STRINGY  WORMHOLES}\\

\vspace{8mm}

 Kamal Kanti Nandi,$^{a,c,}$\footnote{E-mail: kamalnandi@hotmail.com }
 Yuan-Zhong Zhang$^{b,c,}$\footnote{E-mail: yzhang@itp.ac.cn }\\
 and
 K.B.Vijaya Kumar$^{d,}$\footnote{E-mail: kbvijayakumar@yahoo.com}

\vspace{5mm}
  {\footnotesize \it
 $^a$ Department of Mathematics, University of North Bengal,
      Darjeeling (W.B.) 734 430, India\\
 $^b$ CCAST (World Laboratory), P.O.Box 8730, Beijing 100080, China\\
 $^c$ Institute of Theoretical Physics, Chinese Academy of Sciences,
      P.O.Box 2735, Beijing 100080, China\\
 $^d$ Department of Physics, University of Mangalore,
      Mangalore 574 199, India}

\end{center}

\vspace{8mm}

\begin{abstract}
A lower bound on the size of a Lorentzian wormhole can be obtained
by semiclassically introducing the Planck cut-off on the magnitude
of tidal forces (Horowitz-Ross constraint). Also, an upper bound
is provided by the quantum field theoretic constraint in the form
of the Ford-Roman Quantum Inequality for massless minimally
coupled scalar fields. To date, however, exact static solutions
belonging to this scalar field theory have not been worked out to
verify these bounds. To fill this gap, we examine the wormhole
features of two examples from the Einstein frame description of
the vacuum low energy string theory in four dimensions which is
the same as the minimally coupled scalar field theory. Analyses in
this paper support the conclusion of Ford and Roman that wormholes
in this theory can have sizes that are indeed only a few order of
magnitudes larger than the Planck scale. It is shown that the two
types of bounds are also compatible. In the process, we point out
a ``wormhole" analog of naked black holes.

\bigskip

\noindent PACS number(s): 04.20.Gz, 04.50.+h
\end{abstract}

{\bf \vspace{20mm}}

\section{Introduction}
Recent years have seen an intense activity in the field of
wormhole physics especially in the wake of the seminal works of
Morris, Thorne and Yurtsever [1]. Wormholes are created by
embedding into space topological handles that connect two distant
otherwise disconnected regions of space. Theoretical importance of
such geometrical objects is exemplified in several ways. For
instance, they are invoked to interpret/solve many outstanding
issues in the local as well as in cosmological scenarios or even
for probing the interior of black holes [2-5]. Lorentzian
wormholes could be threaded both by quantum and classical matter
fields that violate certain energy conditions (``exotic matter")
at least at the throat. In the quantum regime, several negative
energy density fields are already known to exist. For instance,
they occur in the Casimir effect, and in the context of Hawking
evaporation of black holes, and also in the squeezed vacuum states
[1].  Classical fields playing the role of exotic matter also
exist. They are known to occur in the $R+R^2$ theory [6], scalar
tensor theories [7-11], Visser's cut and paste thin shell
geometries [12]. On general grounds, it has recently been shown
that the amount of exotic matter needed at the wormhole throat can
be made arbitrarily small thereby facilitating an easier
construction of wormholes [13].

A key issue in wormhole physics is the question of traversability.
A wormhole could be traversable in principle but not in practice
due to the occurrence of large tidal forces at and around the
throat. Hence, to ensure the possibility of travel to be realistic
and safe from the human point of view, several classical
constraints are required to be imposed on the parameters of a
Lorentzian wormhole as well as on the kinematics of the traveler.
For instance, the conditions that the time of actual travel be
reasonable and that the tidal accelerations remain less than one
Earth gravity $g_\oplus$ constrain the speed of the traveler in a
definite way. The most severe constraint occurs at the throat of
the wormhole in the form of a radial tension which is inversely
proportional to the square of the throat radius. If the size of
the throat is small, the tension is large. Morris and Thorne [1]
constructed a few wormhole solutions in Einstein's theory and
showed that the velocity of the traveler $v$ is also constrained
linearly by the size $b_0$ of the throat, viz., $v_{th}\leq b_0$
with suitable dimensional adjustments.

In addition to the classical constraints, some of which are
mentioned above, there are constraints that come from the quantum
field theory. For instance, one has the Ford-Roman Quantum
Inequality (FRQI) [14] that provides a constraint of intermediate
nature between pointwise and integral (average) energy conditions.
It has the form of an ``energy density-proper time" quantum
uncertainty type relation that constrains the magnitude and
duration of the negative energy density of a massless minimally
coupled scalar field seen by a timelike geodesic observer. The
validity of these constraints can be illustrated only at the level
of specific, but appropriate, solutions. To this end, Ford and
Roman applied their bound to the stress energy of static,
traversable wormhole spacetimes that were discussed as examples in
Ref.[1]. The calculations demonstrate that the wormholes can only
be microscopic with sizes being a few orders of magnitudes larger
than the Planck scale. Alternatively, if the wormhole is
macroscopic, its geometry must be characterized by large
discrepancy in length scales. Kuhfittig [15] has developed latter
kind of model traversable wormholes by suitably adjusting
different parameters that allow large discrepancies in the
Ford-Roman length scales. However, the solutions considered in the
examples in Ref.[14] were originally designed in Ref.[1] in an
artificial fashion with the primary aim to demonstrate easy
traversability. Not unexpectedly, the resulting stress tensors for
those solutions do not coincide with any {\it a priori} known form
of the source stress tensor provided by some well defined physical
principles. Known forms of stress tensor could come from
physically reasonable theories of gravity such as the minimally
coupled scalar field theory or other field theories mentioned in
the beginning. Apart from this, a desirable feature of any gravity
theory should be that it explains all known tests of gravity to
date. In view of these plausible requirements, we propose to tread
here the reverse path, namely, we start from a premise where the
form of stress energy is known and investigate the semiclassical
and quantum field theoretic constraints on the corresponding
wormhole solutions. We choose to work in the Einstein massless
minimally coupled scalar field (EMS) theory since it is this
theory for which the FRQI was originally intended. To our
knowledge, the literature still seems to lack an investigation of
this kind and the essential motivation of the present paper is to
fill this gap.

In this paper, we shall consider two classes of static,
spherically symmetric exact solutions of the EMS theory which is
just the Einstein frame (EF) version of the low energy limit of
vacuum string theory in four dimensions. That is the reason why we
called such solutions ``stringy" in the title. To go along, the
immediate question to be addressed is whether the considered
solutions truly represent traversable wormholes. This is necessary
in order for any constraint including FRQI to be meaningful. A
detailed analysis shows that, under suitable choices of
parameters, the two classes of solutions do indeed represent
Lorentzian wormholes that are traversable in principle. Practical
traversability, on the other hand,  requires that the magnitude of
tidal forces at the throat be less than the Planck scale. This
condition sets a lower bound (Horowitz-Ross constraint [16]) to
the size of the wormhole throat, which we designate here as a
semiclassical bound in order to distinguish it from the quantum
field theoretic bound. The latter we consider next, namely, the
FRQI and we find, in accord with the conclusions of Ford and
Roman, that the size of the wormholes in the EMS are also bounded
above by values only slightly larger than the Planck scale. Since
both the lower and upper bounds turn out to be of the Planck
order, it is necessary to check that the two bounds are
compatible. It follows that this is also the case. The two
examples that are considered here differ substantially in
character yet, interestingly, they show similar wormhole
behaviors. A couple of limiting cases together with the
interesting wormhole analog of naked black holes are briefly
touched upon. The developments in this paper could be useful also
from the pedagogical point of view.

The paper is organized as follows: In Sec.2, we investigate the
wormhole characteristics of the first example while Sec.3 briefly
touches upon some classical constraints on traversability and the
discussion continues through Sec.4 until we arrive at the
Horowitz-Ross semiclassical constraint. In Sec.5, FRQI is
calculated. In Sec.6, relevant details of the second example are
presented. Sec.7 summarizes the contents. An appendix at the end
contains some useful expressions for the second example.

\section{ EMS theory: Example 1 }
The EMS field equations are given by [9]
 $$R_{\mu \nu }  =  - \alpha  \Phi _{,\mu } \Phi _{,\nu }
                                           \eqno(1)$$
 $$\Phi _{;\mu }^{;\mu }  = 0.            \eqno(2)$$
where $\alpha$ is an arbitrary constant, $\Phi$ is the scalar
field, $R_{\mu\nu}$ is the Ricci tensor and the semicolon denotes
covariant derivatives with respect to the metric $g_{\mu\nu}$. If
$\alpha$ is negative, then the stress tensor of $\Phi$ represents
exotic matter. We shall concentrate here on the solution set given
by ($G = c = \hbar  = 1$):
 $$ds^2  = g_{\mu \nu } dx^\mu  dx^\nu   =  - e^{2\phi (r)} dt^2  +
 e^{ - 2\psi (r)} \left[ {dr^2  + r^2 d\theta ^2  + r^2 \sin ^2
 \theta d\varphi ^2 } \right],              \eqno(3)$$
where
 $$\phi (r) = \psi (r) =  - \frac{M}{r},       \eqno(4)$$
and the scalar field is given by
  $$\Phi (r) =  - \frac{M}{r}.                 \eqno(5)$$
This solution was proposed by Yilmaz [17] decades ago. Integrating
the Einstein complex for the stress energy, we find that the total
conserved mass for the solution is given by $M$ and it is also the
tensor mass that exhibits all the desirable properties of a mass
[18]. Most importantly, the metric (3) exactly coincides up to
second order with the Robertson expansion [19] of a centrally
symmetric field. Hence, it describes all the well known tests of
general relativity just as exactly as does the Schwarzschild
metric for  $r>2M$.

To examine if the solution (3) represents a traversable wormhole
spacetime, it is convenient to employ the five geometric
conditions put forward by Visser [20] which state that: \\
 \indent (i) The functions $\phi (r)$ and $\psi (r)$ are everywhere
 finite. (We call $\phi (r)$ the redshift function).\\
 \indent (ii) The function $C(r) \equiv 2\pi re^{ - \psi (r)}$
 has a minimum at $r_0  \ne 0$. This provides the location of the
 throat at $r=r_0$. \\
 \indent (iii) The two asymptotically flat regions are at
  $r =+\infty$ and at $r=0$. \\
 \indent (iv) $\phi(0)$ and $\phi(\infty)$ must both be finite.\\
 \indent (v) $\psi (\infty)$  must be finite while $e^{ - 2\psi (r)}
 \to r^{ - 4}$ as $r\to 0$.\\
\noindent The condition (i) is obviously satisfied everywhere
except at the origin. The application of the condition (ii) allows
us to locate the wormhole throat at the isotropic coordinate
radius  $r_0 =M$. As for (iii), note that the solution is
asymptotically flat at $r=+\infty$. However, to discover another
flat region at  $r=0$, let us calculate the curvature scalars. The
Ricci, Kretschmann and Weyl scalars respectively turn out to be
 $$R = g^{\mu \nu } R_{\mu \nu }  = \frac{{2M^2 }}{{r^4 }}e^{ -
\frac{{2M}}{r}},                   \eqno(6)$$
 $$R_{\alpha \beta \gamma \delta } R^{\alpha \beta \gamma \delta}=
\left[ {\frac{{28M^4 }}{{r^8 }} - \frac{{64M^3 }}{{r^7 }} +
\frac{{48M^2 }}{{r^6 }}} \right]e^{ - \frac{{4M}}{r}}, \eqno(7)$$
 $$C_{\alpha \beta \gamma \delta } C^{\alpha \beta \gamma \delta}=
 \frac{{16M^2 }}{{3r^8 }}(3r - 2M)^2 e^{ - \frac{{4M}}{r}}.
                     \eqno(8)$$
 All these curvature scalars vanish in the limit $r\to 0$ and so
the spacetime is really flat there. To find the kind of metric
form that exhibits manifest flatness at the origin $r=0$, that is,
a form that satisfies especially the conditions (iv) and (v)
above, we transform the metric (3) under inversion $r\to 1/r$ to
get
 $$ds^2 =- e^{ - 2Mr} dt^2 + r^{ - 4} e^{2Mr} \left[ {dr^2 + r^2
d\theta ^2  + r^2 \sin ^2 \theta d\varphi ^2 } \right].\eqno(9)$$

Now, with regard to condition (iv), note from (9) that $\phi (r)=
e^{ - Mr}  = 1$ at $r=0$ and from (4), $\phi (r) = e^{ -
\frac{M}{r}}  = 1$ at $r=\infty$. Similarly, from (4) again, we
see that $\psi (r) = e^{\frac{M}{r}}  = 1$ at $r=\infty$, while
from (9), it is evident that $e^{ - 2\psi (r)}  \to r^{ - 4}$ as
$r\to 0$ accounting for the condition (v). Thus, finally, we can
conclude that the solution (3) represents a Lorentzian wormhole
that is traversable at least in principle.

The proper radial distance $l$ away from the throat (where $l=0$)
is given by
 $$l(r) =  \pm \int\limits_M^r {e^{\frac{M}{r}} } dr =  \pm \left[
 {re^{\frac{M}{r}}  - M \times Ei\left( {\frac{M}{r}} \right)}
 \right]_M^r,                                  \eqno(10)$$
where $Ei(x)$ is the exponential-integral function given by
 $$Ei(x) = C + \ln x + \sum _{k = 1}^\infty  \frac{{x^k
 }}{{k!k}},\quad \quad x > 0              \eqno(11)$$
 $$\quad \quad = C + \ln ( - x) + \sum _{k = 1}^\infty
 \frac{{x^k }}{{k!k}},\quad \quad x < 0,   \eqno(12)$$
where $C$ is an arbitrary constant. Clearly, it is not possible to
invert Eq.(10) and obtain $r=r(l)$ in a closed form and cast the
metric (3) in the proper distance language.  However, the function
$l$ is well behaved everywhere since $l \to  \pm \infty$ as $r \to
\pm \infty$. The equation for the embedded surface  $z=z(r)$ is
given by
 $$z(r) =\pm\int{\sqrt {\frac{{2M}}{r}-\frac{{M^2 }}{{r^2 }}} }
 \times e^{\frac{M}{r}} dr.                  \eqno(13)$$
Again, the right hand side can not be integrated into a closed
form. Nevertheless, as required, the embedding surface becomes
flat very far from the throat: $dz/dr =0$ as $r \to  \pm \infty$
which corresponds to $l \to  \pm \infty$.

As to the question of violation of energy conditions, note that
the solutions (3)-(5) satisfy the field equations for the value
$\alpha =-2$ [17]. This implies that there is a negative sign
before the kinetic term in the Einstein-Hilbert action.
Consequently, almost all energy conditions are violated providing
a situation that is very conducive for the creation of wormholes.
Indeed, the energy density $\rho$, the radial pressure $p_r$, the
lateral pressures $p_\theta$ and $p_\varphi$ in the static
orthonormal frame turn out to be
 $$\rho  =  - \left( {\frac{1}{{8\pi }}} \right) \times \frac{{M^2
 }}{{r^4 e^{\frac{{2M}}{r}} }},\quad p_r  =  - \left(
 {\frac{1}{{8\pi }}} \right) \times \frac{{M^2 }}{{r^4
 e^{\frac{{2M}}{r}} }},  \quad p_\theta   = p_\varphi
 = \left( {\frac{1}{{8\pi }}}\right) \times \frac{{M^2 }}
 {{r^4 e^{\frac{{2M}}{r}}}}.     \eqno(14)$$
 Clearly, $\rho <0$ for all values of $r$ and the Weak Energy
Condition (WEC) is violated. However, the Strong Energy Condition
(SEC) is marginally satisfied since $\rho  + p_r  + p_\theta   +
p_\varphi   = 0$. The massless limit $M=0$ leads only to a
trivially flat spacetime and is not physically interesting.

\section{Traversability: Classical constraints}

It is of some interest to discuss the classical constraints on
practical traversability across the wormhole by humanoids. To
begin with, note that our wormhole is attractive. The radially
moving traveler that starts off from rest from an asymptotic
location has the equation of motion:
 $$\frac{{d^2 r}}{{d\tau ^2 }} \equiv a^r  = \frac{M}{{r^2 }}
  \times e^{ - \frac{M}{r}}\times\left( {1-\frac{M}{r}} \right),
                                \eqno(15)$$
where $\tau$ is the proper time. Clearly, $a^r  > 0$ for $r>M$ and
$a^r = 0$ for $r=r_0 =M$. Therefore, the traveler will be pulled
in until he/she attains zero acceleration at the throat and in
order to emerge at the other mouth, he/she has to maintain an
outward directed radial acceleration from being pulled in again.
At the throat, the {\it static} observers are also geodesic
observers as $a^r = 0$ there, which is satisfied for a constant
velocity including its zero value [14]. This is a basic feature of
the wormhole example under present investigation.

Suppose that a human being travels radially with velocity $v$ such
that $v=0$ at $l=-l_1$ and at $l=+l_2$ and $v>0$ at $l_1 <l <l_2$,
where $l_1 $ and $l_2$ are the locations of  two widely separated
space stations. Then, in order that the journey is completed in a
reasonable length of time, say, one year, the velocity $v(r)$ has
to satisfy the following constraints [1]:
 $$\Delta \tau  = \int\limits_{ - l_1 }^{ + l_2 }
{\frac{{dl}}{{v\gamma }}}  \le 1{\rm year},\quad \quad \Delta t =
\int\limits_{ - l_1 }^{ + l_2 } {\frac{{dl}}{{ve^\phi  }}}  \le
1{\rm year},                            \eqno(16)$$
 where $\gamma  = \left[ {1 - v^2 } \right]^{ - 1/2}$ and
$\Delta\tau$ is the proper time interval of the journey recorded
by the traveler's clock,  $\Delta t$ is the coordinate time
interval recorded by observers situated at the stations.  These
are also several other kinematic constraints. For instance, at the
stations, the geometry must be nearly flat. This constraint can be
easily satisfied by locating the stations at large $r$. Another
constraint comes from the demand that the traveler not feel an
acceleration greater than one $g_ \oplus$. This leads to
 $$\left| {e^{ - \phi } \frac{{d\left( {\gamma e^\phi  }
\right)}}{{dl}}} \right| \le g_ \oplus.   \eqno(17)$$
 For our solution, the conserved total energy $E$ per unit mass
of the radially freely falling traveler is given by $E = \gamma
(r)e^{\phi (r)} =$constant, and therefore the constraint (17) is
satisfied easily.

\section{Traversability: Horowitz-Ross constraint}

There are also constraints coming from the dynamical
considerations. For instance, traveler's velocity is constrained
by the magnitudes of tidal forces that involve the curvature
tensor. For our form of the solution, the only nonvanishing
curvature components in the static observer's orthonormal basis
are $R_{0101}, R_{0202}, R_{0303}, R_{1212}, R_{1313}$ and
$R_{2323}$. Radially freely falling travelers with conserved
energy $E$ per unit mass are connected to the static orthonormal
frame by a local Lorentz boost with an instantaneous velocity
given by
 $$v = \frac{{dr}}{{d\tau }} = \left[ {1 - e^{2\phi } E^{ - 2} }
\right]^{1/2}.                      \eqno(18)$$
 Then the nonvanishing curvature components in the Lorentz-boosted
 frame ($\land$) are [16,21]:
 $$R_{\hat 0\hat 1\hat 0\hat 1}  = R_{0101},      \eqno(19)$$
 $$R_{\hat 0k\hat 1k}  = \cosh \alpha \sinh \alpha \left( {R_{0k0k}
+ R_{1k1k} } \right),                            \eqno(20)$$
 $$R_{\hat 0k\hat 0k}= R_{0k0k} +\sinh ^2 \alpha \left( {R_{0k0k}
+ R_{1k1k} } \right),                             \eqno(21)$$
 $$R_{\hat 1k\hat 1k}= R_{1k1k}+ \sinh ^2 \alpha \left( {R_{0k0k}
+ R_{1k1k} } \right),                             \eqno(22)$$
 and $R_{klkl}$, where $k, l=2,3$ and $\sinh \alpha  = \frac{v}
 {{\sqrt {1 - v^2 } }}$. The terms in the parentheses represent an
enhancement of curvature in the traveler's frame. Incidentally,
note that, in the Schwarzschild or Reissner-Nordstr\"{o}m
spacetime, the sums in the parentheses are exactly zero due to
special cancellations. This might appear surprising at first
sight, but actually this cancellation occurs only in the
``standard" coordinates which hide the nontrivial enhancement that
{\it actually} takes place. This is only to be expected as the two
pieces in $\left( {R_{0k0k}  + R_{1k1k} } \right)$ transform
differently at any spacetime point under transformations to
different coordinate systems.\\

The differential tidal accelerations felt by the traveler are
 $$\Delta a_j  =  - R_{\hat 0j\hat 0p} \xi ^p,        \eqno(23)$$
where  $j,p=1,2,3$ and $\vec \xi$ is the vector separation between
two parts of the body. Taking $\left| {\vec \xi } \right| \approx
2{\rm meters}$ (the size of the body), the radial tidal constraint
should be such as to satisfy $\left| {R_{\hat 0\hat 1\hat 0\hat 1}
} \right| \le \frac{{g_ \oplus  }}{{2m}} \cong 10^{ - 20} cm^{ -
2}$.  For the solution (4), we have:
 $$\left| {R_{\hat 0\hat 1\hat 0\hat 1}}\right| =\left| {R_{0101}
} \right| = \frac{{2Me^{ - \frac{{2M}}{r}} }}{{r^3 }}\left( {1 -
\frac{M}{r}} \right),                               \eqno(24)$$
which vanishes at the throat $r=r_0 =M$. Evidently, the constraint
is well satisfied throughout the journey. On the other hand, the
requirement
 $$\left| {R_{\hat 02\hat 02} } \right| \le \frac{{g_ \oplus
}}{{2m}} \cong 10^{ - 20} cm^{ - 2}                    \eqno(25)$$
constrains the velocity $v$ of the traveler to values that are
comfortably attainable[1].  The exact form of $\left| {R_{\hat
02\hat 02} } \right|$ will be shown below. However, from now on,
we shall focus on the constraints engendered by physical
requirements rather than by the requirement of human comfort.
Using Eq. (21), we have
 $$\left| {R_{\hat 02\hat 02}} \right| = \frac{{Me^{ -
\frac{{2M}}{r}} }}{{r^3 }}\left( {1 - \frac{M}{r}} \right) +
\frac{{M^2 e^{-\frac{{2M}}{r}}v^2 \gamma ^2}}{{r^4 }}.\eqno(26)$$
Clearly, the first term on the right is the curvature measured in
the static frame while the second represents excess in curvature
measured by the geodesically falling observer with $v \ne 0$.
Other curvature components follow from Eqs. (20) and (22) and they
are:
 $$\left| {R_{\hat 03\hat 03} } \right| = \frac{{Me^{ -
\frac{{2M}}{r}} }}{{r^3 }}\left( {1 - \frac{M}{r}} \right) +
\frac{{M^2 e^{ - \frac{{2M}}{r}} v^2 \gamma ^2 }}{{r^4 }},\quad
\left| {R_{\hat 02\hat 12} } \right| = \frac{{M^2 e^{ -
\frac{{2M}}{r}} v\gamma ^2 }}{{r^4 }},     \eqno(27)$$
 $$\left|{R_{\hat 12\hat 12}}\right| \equiv \left| {R_{\hat 13\hat
13} } \right| = \frac{{Me^{ - \frac{{2M}}{r}} }}{{r^3 }}\left( {1
+ \frac{{Mv^2 \gamma ^2 }}{r}} \right),\quad \left| {R_{\hat 2\hat
3\hat 2\hat 3} } \right| \equiv \left| {R_{2323} } \right| =
\frac{{Me^{ - \frac{{2M}}{r}} }}{{r^2 }}\left( {\frac{M}{{r^2 }} -
\frac{2}{r}} \right),                               \eqno(28)$$
and they can also be separated likewise into static and excess
parts. It may be noted here that, at the throat, $r=r_0 =M$, all
the values of the curvature remarkably coincide, up to an
unimportant factor ($e^2 /2$), with those obtained for the case of
``$\phi  = 0, \; b = r_0  = {\rm const.}$" zero density wormholes
discussed in Refs.[1, 14].

For a particle that is static at the throat, the radial and
lateral tidal forces, given respectively by $\left| {R_{0101}}
\right|$, $\left| {R_{0202}} \right|$ and $\left| {R_{0303}}
\right|$, are exactly zero. But a radially falling particle could
experience much larger tidal forces  in the vicinity of the
throat,  either for its velocity $v \approx 1$ or for the wormhole
geometry $r_0  \approx 0$ or for both reasons, than the one static
at the throat that actually feels no tidal forces at all. Thus, we
have here a wormhole analog of the idea of naked black holes
proposed by Horowitz and Ross [16] for which the curvatures just
above the horizon are much larger than those at the horizon. In
the vicinity of the throat, the maximum value of the curvature
felt by the falling particle [Eqs.(26)-(28)] is given by $\gamma
_0^2 /r_0^2$, where $\gamma _0 $ is the Lorentz factor at the
throat. In order to avoid the occurrence of infinite tidal forces,
the physical requirement is that the magnitude of curvature be
less than the Planck scale. This implies that the local radius of
curvature ($r_0 /\gamma_0$) be greater than the Planck length.
This condition gives us a semiclassical lower bound or a Planck
cut-off, on $r_0$, and we call it the Horowitz-Ross constraint
[16], viz.,
 $$r_0  > \gamma _0 \ell _P,                          \eqno(29)$$
where $\ell _P$ is the Planck length, $\gamma _0 =1/\sqrt {1 -
v_0^2 }$ and $v_0$ is the velocity of the particle at the throat.
Due to the introduction of the Planck length, the right hand side
of (29) remains microscopic even for values of $v_0$ very close to
unity. Inequalities similar to (29), but without involving the
Planck scale, have also been worked out by Morris and Thorne [1]
in case of their examples of traversable wormholes. We shall now
turn to FRQI to see what upper bound it offers on the throat size.

\section{Ford-Roman Quantum Inequality (FRQI)}
This is a constraint coming essentially from the full quantum
field theoretic considerations. The bound has the form of an
uncertainty-principle-type constraint on the magnitude and
duration of the negative energy density as seen by an observer
fixed to a timelike geodesic particle. The quantum inequality is
given by [14]:
 $$\frac{{\tau _0 }}{\pi }\int\limits_{ - \infty }^{ + \infty }
{\frac{{\left\langle {T_{\mu \nu } u^\mu  u^\nu  } \right\rangle
d\tau }}{{\tau ^2  + \tau _0^2 }}}  \ge  - \frac{3}{{32\pi ^2 \tau
_0^4 }},                                             \eqno(30)$$
for all $\tau _0$ where $\tau$ is the freely falling observer's
proper time, $\left\langle {T_{\mu \nu } u^\mu  u^\nu  }
\right\rangle$ is the expectation value of the stress energy of
the minimally coupled scalar field in the observer's frame of
reference. Although the inequality was basically derived in the
Minkowski space quantum field theory, it can be applied also in
the curved spacetime provided that $\tau_0$ is taken sufficiently
small, that is, much less than the size of the proper local radius
of curvature.

To apply the FRQI to our solution, let us find the energy density
in the geodesic frame of the radially falling observer. This can
be obtained by applying a local Lorentz boost given by
 $$\rho '=\gamma^2 \left({\rho + v^2 p_r}\right).    \eqno(31)$$
Using the relevant expressions from Eqs.(14), we have
 $$\rho ' =-\left( {\frac{1}{{8\pi }}} \right) \times \frac{{M^2
\gamma ^2 }}{{r^4 e^{\frac{{2M}}{r}} }} \times \left( {1 + v^2 }
\right) < 0.                                        \eqno(32)$$
Next, from the expressions of the components of Riemann tensor
[Eqs.(26)-(28)], it follows that, at the throat, the maximum
magnitude of curvature in the Lorentz-boosted frame is $R'_{\max }
\le \frac{{\gamma _0^2 }}{{r_0^2 }}$ and therefore the smallest
local proper radius of curvature $r'_c  \ge r_0 /\gamma _0$. Thus
the sampling time is taken as $\tau _0  = fr_0 /\gamma _0 <<
r'_c$, for $f << 1$. The energy density does not significantly
change over this time scale and FRQI says:
 $$\frac{{\tau _0 }}{\pi }\int\limits_{ - \infty }^{ + \infty }
 {\frac{{\left\langle {T_{\mu \nu } u^\mu  u^\nu  } \right\rangle
 d\tau }}{{\tau ^2  + \tau _0^2 }}}  \approx \rho '_0  \ge  -
 \frac{3}{{32\pi ^2 \tau _0^4 }},      \eqno(33)$$
where $\rho '_0  =  - \left( {\frac{1}{{8\pi }}} \right) \times
 \frac{{\gamma _0^2 }}{{r_0^2 e^2 }} \times \left( {1 + v_0^2 }
 \right)$ is the value of the energy density at the throat
$r=r_0 =M$. Putting this value in FRQI (33), we have the upper
bound on $r_0$:
 $$r_0  \le \left( {\frac{e}{{2f^2 \sqrt {1 - v_0^4 } }}}
 \right)\ell_P.                     \eqno(34)$$

For $v_0  = 0$ (recall that it is still geodesic motion), and $f
\approx 10^{ - 4}$, we have $r_0 \approx 10^{ - 29}$cm. Even if
$v_0$ is extremely close to unity, say $1 - v_0^4  \approx 10^{ -
40}$, one has $r_0 \approx 10^{ - 5}$cm. These results show that
the FRQI bound is really robust. The solution (3) of the EMS
theory does indeed represent a wormhole of microscopic size, even
at the two near extreme values of observer's velocity. Considering
a realistic motion (with energy $E$ normalized to unity) that
begins from rest at the asymptotic region and passes through the
wormhole throat, we see that the particle attains maximum velocity
right at the throat and it is $v = v_0  = \sqrt {1 - e^{ - 2} }$.
Then $r_0 \le (1/2f^2 )(e^3 /\sqrt {2e^2  - 1} )\ell _P$.
Obviously, again the FRQI constrains the wormholes to have sizes
that are just a few orders of magnitude larger than the Planck
scale. Looking at the Horowitz-Ross constraint (29), we expect it
to be compatible with the FRQI (34). That is, we expect the
following inequality to hold good:
 $$\gamma _0 \ell _P  < r_0  \le \left( {\frac{e}{{2f^2 \sqrt {1 -
 v_0^4 } }}} \right)\ell _P .                 \eqno(35)$$
This is true if $\sqrt {1 + v_0^2 }  < \frac{{e^2 }}{{2f^2 }}$,
which is easily satisfied for $f<<1$.

It was mentioned earlier that the class of solutions (4) is
distinguished from other classes of solutions in the EMS theory in
some important respects. In the next section, we consider one such
class of solutions in the form of a second example pointing out
how it differs in nature from that of Example 1.

\section{EMS theory: Example 2}
Consider the class of solutions, which, in isotropic coordinates,
is given by [9,22]:
 $$\phi (r) = \beta \ln \left[ {\frac{{1 - \frac{m}{{2r}}}}{{1 +
 \frac{m}{{2r}}}}} \right],\quad \quad \psi (r) = (\beta  - 1)\ln
 \left( {1 - \frac{m}{{2r}}} \right) - (\beta  + 1)\ln \left( {1 +
 \frac{m}{{2r}}} \right),                \eqno(36)$$
 $$\Phi (r) = \left[ {\frac{{2\left( {1 - \beta ^2 }
 \right)}}{\alpha }} \right]^{\frac{1}{2}} \ln \left[ {\frac{{1 -
 \frac{m}{{2r}}}}{{1 + \frac{m}{{2r}}}}} \right], \eqno(37)$$
where $\alpha$ is an arbitrary constant parameter. The two
undetermined constants $m$ and $\beta$ are related to the source
strengths of the gravitational and scalar parts of the
configuration. To highlight the differences in nature between this
solution and that in (4), we point out the following: Once the
scalar component is set to zero ($\Phi  = 0 \Rightarrow \beta =
1$), the solutions (36), (37) reduce to the Schwarzschild black
hole in accordance with Wheeler's ``no scalar hair" conjecture.
Physically, this indicates the possibility that the scalar field
is radiated away during collapse and the end result is a
Schwarzschild black hole [18]. On the other hand, in the case of
our previous example, solution (4), there is no separate scalar
parameter. The condition $\Phi  = 0 \Rightarrow M = 0$, that is,
one obtains only a flat space from the metric (3) and not a black
hole. In this sense, (4) was a pure wormhole solution having no
counterpart in the black hole regime. Another important difference
is that, for $\beta\ne 1$, the solutions (36), (37) represent a
spacetime with naked singularity at $r=m/2$ in the sense that all
curvature invariants diverge there. In contrast, such divergences
do not occur in the solutions (4), (5). In spite of these basic
differences, the calculations below show that the presence of a
separate scalar parameter $\beta$ does not alter the Horowitz-Ross
or FRQI constraints.

The solution (36) can be interpreted as a traversable wormhole as
it satisfies all of the Visser's conditions (i)-(v). We only
mention that the metric is not only flat at $r=0$ but is also form
invariant under inversion. [Just choose $r = (m/2)\rho ,\quad \rho
\to 1/\rho$.] The throat appears at the coordinate radii
 $$r_0^ \pm = \frac{m}{2}\left[ {\beta \pm \left( {\beta ^2 - 1}
 \right)^{1/2} } \right].                         \eqno(38)$$
Here we take only the positive sign ($r_0 ^+$). The requirement
that the throat radii be real implies that $\beta^2 >1$ and the
reality of $\Phi$ in turn demands that $\alpha <0$. Alternatively,
one could have $\alpha >0$ allowing for an imaginary $\Phi$. The
latter choice presents no pathology or inconsistency in the
wormhole physics, as recently shown in Ref. [23]. In both cases,
however, we have a negative sign before the stress tensor on the
right hand side of Eq. (1) and consequently almost all energy
conditions are violated. For instance, the energy density is given
by
 $$\rho  =  - \left( {\frac{1}{{8\pi }}} \right) \times \left[
{\frac{{256m^2 r^4 (\beta ^2  - 1)(1 - m/2r)^{2\beta } (1 +
 m/2r)^{ - 2\beta } }}{{(m^2  - 4r^2 )^4 }}} \right].\eqno(39)$$
Thus, $\rho <0$ at the throat and elsewhere satisfying the
necessary wormhole condition that the Weak Energy Condition (WEC)
be violated. The expressions for the pressure components are given
in the Appendix. Note that the tensor mass of the solution is
given by $M=m\beta$ and the expansion of the metric (36) indicates
that it is also the Keplerian mass.

All curvature components in the Lorentz boosted orthonormal frame
are given in the Appendix. We consider here only a representative
one, viz., $R_{\hat 02\hat 02}$. From (A4)-(A6), it is evident
that the static frame measure of the curvature at the throat is
zero. The geodesic excess, at the throat $r=r_0^+ (>m/2)$ is given
by the last term in (A5) which works out to a remarkably simple
expression:
 $$\left| {R_{\hat 02\hat 02}} \right| = \left( {\frac{{v_0
 \gamma _0 }}{{r_0^ +  }}} \right)^2 ,          \eqno(40)$$
for $\beta^2 >1$. So, once again, we get the same Horowitz-Ross
constraint, that is, the inequality (29). Note that, for $\beta
=1$ (Schwarzschild), we have, $r_0^ +   \equiv r_H  = \frac{m}{2}$
($r_H$ is the horizon radius) and $v_0 =1$, as expected. Only for
these exact values, $\left| {R_{\hat 02\hat 02}} \right| \to
\infty$, that is, an arbitrarily large tidal force is experienced
by the test (light!) particle. But for slightly massive test
particles ($v_0  \approx 1$), one can introduce a Planck cut-off
as embodied in (29) and avoid infinities in the measurement of
curvature.

As to the FRQI, we get, at the throat, using a little manipulation
with Eqs. (31), (39) and (A1),
 $$\rho ' =  - \frac{1}{{32\pi }} \times \frac{{\gamma _0^2 (1 +
 v_0^2 )}}{{{r_{0}^{+}}^2 }} \times \left( {\frac{{\beta  + \sqrt
  {\beta ^2  - 1} }}{{\sqrt {\beta ^2  - 1} }}} \right)^2  \times
 \left[ {\frac{{(\beta  - 1) - \sqrt {\beta ^2  - 1} }}{{(\beta  -
 1) + \sqrt {\beta ^2  - 1} }}} \right]^{2\beta }$$
 $$\equiv -\frac{1}{{32\pi }}\times\frac{{\gamma _0^2 (1+v_0^2 )}}
 {{{r_0^+}^2 }} \times g(\beta ).                  \eqno(41)$$
For $\beta\to 1$ it is clear from (39) and (A1) that both $\rho$
and $p_r$ vanish implying that $\rho ' = 0$. For $\beta\to\infty$,
the function $g(\beta)$ tends to 0. Therefore, only the
coefficient of $g(\beta)$ is important for FRQI. Noting from
Eq.(40), that $R'_{\max }  \le {\gamma _0^2 }/{{r_0^+}^2 }$ and
taking $\tau _0 = fr_0^ +  /\gamma _0$ with $f<< 1$, and using Eq.
(41) in (33), we get the same upper bound as in (34). This
concludes the discussion of bounds on wormholes.

Let us consider a couple of limiting cases. If $m \approx 0$, one
can choose $\beta$ sufficiently large and arrange to have any
finite nonzero value for $M$ so that
 $$r_0^ +   \approx M,\quad \quad \left. \rho  \right|_{r_0^ +  }
 \approx -\frac{1}{{8\pi M^2 }},\quad \quad \left| {R_{\hat 02\hat
 02}} \right| = \frac{{v_0^2 \gamma _0^2 }}{{M^2 }},  \eqno(42)$$
with all other curvature components behaving similarly. This is
the closest approximation to a Schwarzschild-like, but traversable
wormhole that that one can obtain in the EMS theory.  If, on the
other hand, we set $\beta =0$ but $m\ne 0$ in the equation
$M=m\beta$, we have a zero mass ($M=0$) wormhole [23]. These
solutions are not flat. In fact, in this case, we have
 $$r_0^ + = \frac{{m'}}{2},\quad\quad \left. \rho \right|_{r_0^ +
 }=  - \frac{1}{{8\pi m'^2 }},\quad \quad {\bf R} =- \frac{2}{{m'^2
 }},\quad \quad m =  - im',          \eqno(43)$$
where ${\rm\bf R}$ is the Ricci scalar. This is an extreme case,
since $M=0$. Also, at the throat the velocity of the test particle
for unit energy $E$, viz.,
 $$v_0^2 =1 -\left[ {\frac{{\left( {\beta - 1} \right) - \sqrt
 {\beta ^2 - 1} }}{{\left( {\beta - 1} \right)+ \sqrt {\beta ^2
 - 1} }}} \right]^{2\beta }               \eqno(44)$$
becomes zero for $\beta =0$. (This also implies that $\left|
{R_{\hat 02\hat 02}} \right| = 0$.) Therefore, the test particle
is captured and kept at rest forever at the throat [24, 25]. This
is an interesting aspect of zero mass wormholes.

\section{Summary}
Quantum field theory calculations involving massless minimally
coupled scalar field (EMS theory) imply that there are two
possible alternatives: Either a wormhole threaded by this matter
must only be of microscopic (Planck) size or that there should be
large discrepancies in the length scales associated with
macroscopic wormholes [14]. Ford and Roman applied their bound
only to some artificial examples for which the stress tensors do
not comply with those in the EMS theory. There is therefore the
important logical need that the bound be applied in the proper
setting. To this end, it is necessary to consider exact wormhole
solutions in the EMS theory, investigate their traversability and
see which of the two alternatives is allowed. The present paper is
motivated essentially by these considerations.

We considered two wormhole examples from the EMS theory. The first
example has been worked out in some detail while analogous
calculations can be carried out for the second example, of which
an outline is given above. In both the examples, we calculated the
physical condition for traversability which provides the
Horowitz-Ross lower bound [16] on the throat size of the wormhole.
This bound is obtained by introducing the Planck cut-off on a
classical quantity, viz., curvature and that is why we called this
bound semiclassical. In the process, we arrived at the wormhole
analog of naked black holes proposed by Horowitz and Ross [16].
The similarity is interesting given that the energy conditions are
violated only in the former case, but not in the latter. The FRQI
provides a quantum field theoretic upper bound on the throat size.
It is shown that the two bounds are compatible. The main lesson
that the two examples teach us is that traversable Lorentzian
wormholes in the EMS theory could indeed be microscopic, which
supports the conclusions of Ford and Roman [14] in a direct way.
An analogous result has been advanced by Visser [26] in the
context of minisuperspace models. He has shown that the
expectation value of the throat radius is also of the order of
Planck length. It is tempting to speculate that the EMS wormholes,
in virtue of their sizes being microscopic, could be the {\it
natural} candidates for the constituents of the spacetime ``foam"
of Wheeler [27, 28].

Finally, although microscopic wormholes are of considerable
theoretical interest, one question still remains. Recall that
traversability is a basic criterion in order for FRQI to be
defined since the negative energy density is measured in the
proper frame of the traveling or static observer. If the wormhole
throats are doomed to be of only Planck dimensions in the EMS
theory, can one meaningfully define a {\it non-hypothetical}
static and/or a traveling test particle through the wormhole? It
seems, in general, one can't since the Bohr radius of an
elementary particle is several orders of magnitude higher than the
Planck length. However, if the velocity is exceedingly close to
that of light, an elementary particle can just pass through [see
the discussion after Eq.(34)]. For zero mass wormholes, the test
particle is captured at the throat and kept at rest forever there.
The possibility of interstellar travel by using these microscopic
objects seems out of question [26].

\section*{Acknowledgments}
One of us (KKN) wishes to thank Professor Ouyang Zhong Can for
providing hospitality and excellent working facilities at ITP,
CAS. Administrative assistance from Sun Liqun at ITP of CAS is
gratefully acknowledged. This work is supported in part by the
TWAS-UNESCO program of ICTP, Italy and the Chinese Academy of
Sciences, well as in part by National Basic Research Program of
China under Grant No. 2003CB716300 and by NNSFC under Grant
No.10175070.

\section*{Appendix}
For the metric (36), the pressure components are given by
 $$p_r  =  - \left( {\frac{1}{{8\pi }}} \right) \times \left[
{\frac{{256m^2 r^4 (\beta ^2  - 1)(1 - m/2r)^{2\beta } (1 +
 m/2r)^{ - 2\beta } }}{{(m^2  - 4r^2 )^4 }}} \right],\eqno(A1)$$
 $$p_\theta   = p_\varphi   =  - p_r .        \eqno(A2)$$
Using the expression for $\rho$ from Eq.(39), we have
 $$\rho  + p_r  + p_\theta   + p_\phi   = 0. \eqno(A3)$$

The curvature components in the Lorentz boosted orthonormal frame
for the metric (36) read, using Eqs. (19)-(22):
 $$R_{\hat 01\hat 01}  = R_{0101}  = \frac{{128m\beta r^3 (m^2  +
 4r^2  - 4m\beta r)(1 - m/2r)^{2\beta } (1 + m/2r)^{ - 2\beta }
 }}{{(m^2  - 4r^2 )^4 }},                    \eqno(A4) $$
 $$R_{\hat 02\hat 02}= R_{\hat 03\hat 03}= R_{0202}+ v^2 \gamma
 ^2 (R_{0202}  + R_{1212} ),                 \eqno(A5)$$
 $$R_{\hat 12\hat 12}  = R_{1212}  + v^2 \gamma ^2 (R_{0202}  +
 R_{1212} ),                                \eqno(A6)$$
 $$R_{\hat 2\hat 3\hat 2\hat 3}= R_{2323}= \frac{{128mr^3 [m^2
 \beta  + 4r^2 \beta - 2mr(1+ \beta ^2 )](1 - m/2r)^{2\beta }(1
 + m/2r)^{ - 2\beta } }}{{(m^2  - 4r^2 )^4 }},   \eqno(A7)$$
 $$R_{\hat 02\hat 12}  = v\gamma ^2 (R_{0202}  + R_{1212} ),
                                              \eqno(A8)$$
 $$R_{0202}  = R_{0303}  = \frac{{64m\beta r^3 (m^2  + 4r^2  -
 4m\beta r)(1- m/2r)^{2\beta }(1 + m/2r)^{- 2\beta }}}{{(m^2 -
 4r^2 )^4 }},                                \eqno(A9)$$
 $$R_{1212}=R_{1313}= \frac{{128mr^3 (m^2 \beta  - 4mr + 4r^2
 \beta )(1- m/2r)^{2\beta }(1 + m/2r)^{- 2\beta } }}{{(m^2  -
 4r^2 )^4 }}.                              \eqno(A10)$$
The wormhole throat satisfies ${r_0^\pm}^2  + \frac{{m^2 }}{4} -
m\beta r_0^ \pm   = 0$, and so, from (A4) and (A5), we see that in
the static frame, the tidal accelerations [Eq. (23)] vanish at the
throat.

\end{document}